\documentclass[%
 prreseach,
 cha,
 amsmath,amssymb,
 reprint,%
twocolumn,showpacs,amsmath,amssymb,superscriptaddress,longbibliography]{revtex4-2}
\usepackage{float}
\usepackage{graphicx}% Include figure files
\usepackage{dcolumn}% Align table columns on decimal point
\usepackage{bm}% bold math
\usepackage{siunitx}
\DeclareSIUnit\gauss{G}
\usepackage{enumerate}
\usepackage{multirow}
\usepackage{xcolor}
\usepackage{enumerate}
\usepackage{hyphenat}
\usepackage{notoccite}
\makeatletter
\newcommand{\ORCID}[1]{\href{https://orcid.org/#1}{\includegraphics[height=\f@size pt]{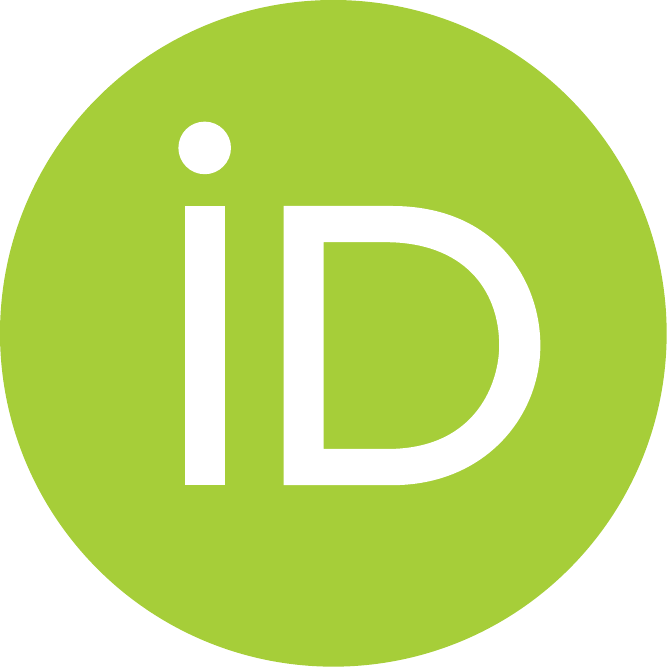}}}
\makeatletter

\newcommand{\ket}[1]{\ensuremath{\left|#1\right>}}
\newcommand{\figref}[1]{Fig.~\ref{#1}}

\newcommand{\eref}[1]{Eq.~(\ref{#1})}

\newcommand{\fref}[1]{Fig.~\ref{#1}}
\newcommand{\tref}[1]{Table~\ref{#1}}

\newcommand{\Fref}[1]{Figure~\ref{#1}}

\DeclareMathOperator{\real}{Re}
\DeclareMathOperator{\imag}{Im}
\usepackage{xcolor}
\definecolor{roed}{RGB}{0,100,180}
\usepackage[colorlinks=true,urlcolor=roed,citecolor=roed,linkcolor=roed]{hyperref}
\begin{document}
\preprint{AIP/123-QED}

\title{Experimental observation of the avoided crossing of two $S$-matrix
	resonance poles in an ultracold atom collider}
\author{Matthew Chilcott\ORCID{0000-0002-1664-6477}}
\affiliation{ 
	Department of Physics, QSO---Quantum Science Otago, and Dodd-Walls
	Centre for Photonic and Quantum Technologies, University of Otago, New Zealand
}%
\author{Ryan Thomas\ORCID{0000-0002-4308-8781}}%
\affiliation{ 
	Department of Physics, QSO---Quantum Science Otago, and Dodd-Walls
	Centre for Photonic and Quantum Technologies, University of Otago, New Zealand
}%
\affiliation{Department of Quantum Science, Research School of Physics, The Australian National University, Canberra 2601, Australia}

\author{Niels Kj{\ae}rgaard\ORCID{0000-0002-7830-9468}}%
\email{niels.kjaergaard@otago.ac.nz}
\affiliation{ 
	Department of Physics, QSO---Quantum Science Otago, and Dodd-Walls
	Centre for Photonic and Quantum Technologies, University of Otago, New Zealand
}%
\date{\today}% It is always \today, today,
             %  but any date may be explicitly specified

\begin{abstract}
In quantum mechanics, collisions between two particles are captured by a scattering matrix which describes the transfer from an initial entrance state to an outgoing final state.
Analyticity of the elements of this $S$-matrix enables their continuation onto the
complex energy plane and opens up a powerful and widely used framework in scattering theory, where bound states and
scattering resonances for a physical system are ascribed to $S$-matrix poles.
In the Gedankenexperiment of gradually changing the potential parameters of the system, the complex energy
poles will begin to move, and in their ensuing flow, two poles approaching will interact.
An actual observation of this intriguing
interaction between scattering poles in a collision experiment has,
however,  been elusive. Here, we expose the interplay between
two scattering poles relating to a shape resonance and a magnetically
tunable Feshbach resonance by studying ultracold atoms with a laser-based collider. We exploit the tunability of the Feshbach
resonance to observe a compelling avoided crossing of the poles in
their energies which is the hallmark of a strongly coupled system.
\end{abstract}

\maketitle

\section{Introduction}
Atomic collision resonances occur when a scattering
state in the energy continuum couples to a quasi-bound state of the
inter-atomic potential. Near a resonant energy, the scattering of
colliding atoms may be dramatically affected and either enhanced or
suppressed. According to the superposition principle of quantum
mechanics, the final state of the system results from a linear
operator $\hat{S}$ acting on the initial state of the incoming
particles \cite{Eden1966}. The $S$-matrix responsible for this
transformation is a function of the centre-of-mass collision energy
$E$, which is a positive, real quantity for a physical process. It
may, however, be analytically continued into the complex plane for
most physical interaction potentials. Remarkably, poles of this
analytically extended $S$-matrix occurring at nonphysical, complex
values of $E$ are intimately related to the scattering resonances
observed when scanning the collision energy along the real energy axis
\cite{taylorScattering,Sitenko1971,Kukulin1989}. Indeed,
characterizing a resonance through its pole position provides a
powerful alternative to Breit-Wigner parametrization \cite{Ceci2017,Zyla2020}.

The locations of $S$-matrix poles are determined by the
interaction potential of the system Hamiltonian.  As a consequence, upon
changing the parameters describing the potential the poles will move and their
ensuing flow has been investigated theoretically, for example, in the
cases of various potential wells
\cite{Nussenzveig1959,Potvliege1988,Dabrowski1997,Belchev2011,Racz2011,Meeten2019},
or coupled two-state systems \cite{Vanroose1997,Rotter2001,
	Kokkelmans2002,Rakityansky2006,Klosiewicz2012}. In the latter
instance it was shown that inter-channel coupling may give rise to
avoided crossings of poles similarly to the Landau-Zener avoided
crossing for coupled discrete states of a bound system. An observation
of this intriguing interplay between poles of the $S$-matrix in a
collision experiment has, however, been lacking.

In this study, we experimentally investigate atomic scattering when
both a shape resonance and a Feshbach resonance are at work at the
same time. In particular, we observe the signature of an avoided
crossing between the two $S$-matrix poles corresponding to the
resonances. We consider rubidium atoms colliding at energies in the
nano-eV range, with one pole originating from a $d$-wave shape
resonance and the other from a $d$-wave magnetic Feshbach
resonance. Our system is well described by a two-channel picture and
illuminates with remarkable clarity the interplay between the two
archetypal types of resonances found in atomic scattering: shape and
Feshbach resonances. Because both resonances have a $d$-wave character,
the interplay dramatically affects the scattering cross section at the
nominal position of the shape resonance while scattering at threshold
remains s-wave and largely unaffected.
\section{Scattering resonances} In atomic scattering,
shape resonances are typically formed by coupling to a quasi-bound state
accommodated behind the centrifugal barrier for collisions with
non-zero angular momentum. Atomic shape resonances have been observed
using a number of techniques including photoassociation spectroscopy
\cite{PhysRevA.55.636}, molecule loss spectroscopy \cite{Yao2019}, molecular dissociation
\cite{Volz2005}, or in collider-like settings \cite{PhysRevLett.81.5780,
	PhysRevLett.93.173201, PhysRevLett.93.173202,Thomas2016}. Residing in the
open-channel the quasi-bound
state, and hence the resonance, typically only has minuscule
tuning with external fields \cite{Yao2019}. \nocite{dummy}\nocite{Tiesinga1996,Mies2000}
\begin{figure*}[t!]
	\includegraphics[width=\linewidth]{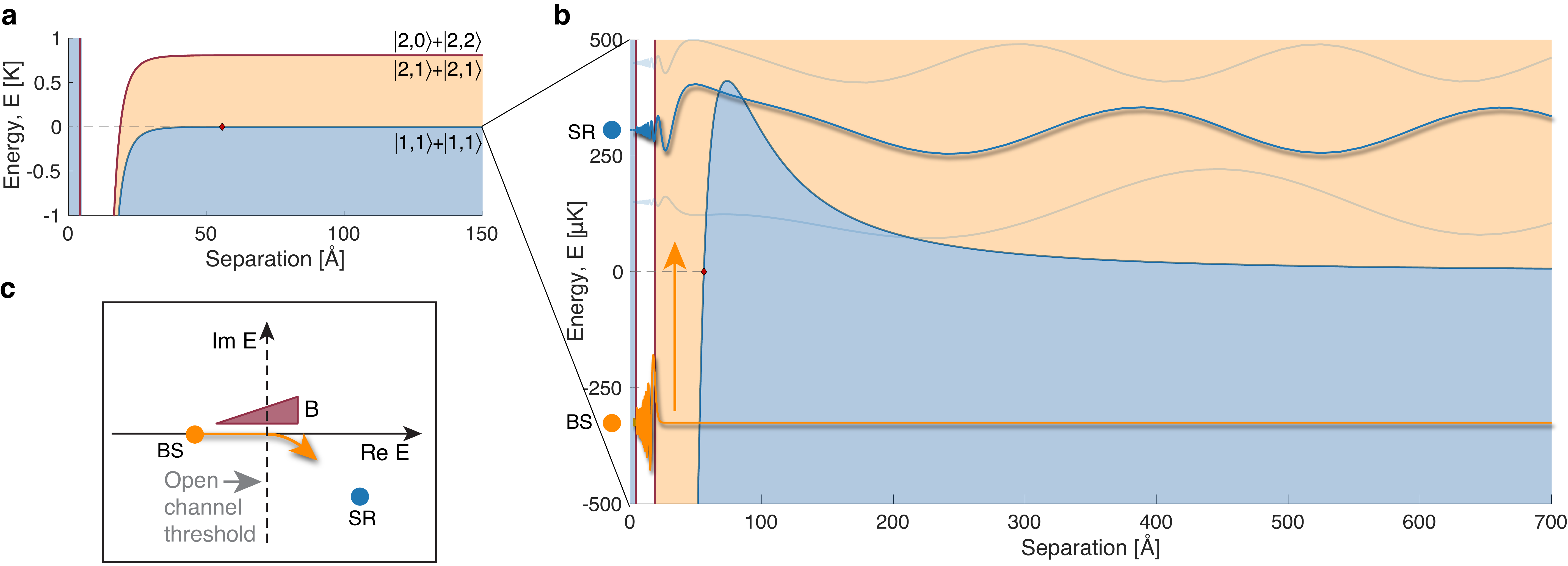}
	\caption{{Scenario for atomic scattering with two $S$-matrix poles simultaneously in play.} {(a)} Relevant atomic-basis potentials for cold $d$-wave collisions between two $\rm^{87}Rb$ atoms in the $|F=1,m_F=1\rangle$ hyperfine state \cite{dummy}. The difference between the $|2,0\rangle+|2,2\rangle$ and $|2,1\rangle+|2,1\rangle$ potentials is not discernible for the window shown. The energy scale also renders the $\ell=2$ centrifugal barriers invisible, but the red diamond shows the classical turning point inside the barrier for the $|1,1\rangle+|1,1\rangle$ channel at threshold (dashed line). {(b)}
		Scattering state wave functions in the $|1,1\rangle+|1,1\rangle$ entrance channel
		[blue, with dark blue highlighting a $d$-wave shape
		resonance (SR)]. For an external magnetic field $B=927.5$~
		G a bound state (BS) resides below threshold of the open channel: the orange wave function shows its $|2,0\rangle+|2,2\rangle$ component; the other significant component (not shown), $|2,1\rangle+|2,1\rangle$, is very similar. Upon increasing the external magnetic field, the position of the bound state moves up in energy as indicated by the arrow. It eventually crosses the $|1,1\rangle+|1,1\rangle$ threshold to become a quasi-bound state corresponding to a Feshbach resonance in this channel. (c) The bound state and the shape resonance correspond to $S$-matrix poles on the complex energy plane and the magnetic field will tune their relative positions.
	}
	\label{fig:wavefunction}
\end{figure*}

In contrast to shape resonances, Feshbach resonances are highly tunable. These occur due
to the presence of a bound state in an energetically closed channel
which has a different spin to the entrance channel. With inter-channel coupling present
one can use the Feshbach resonance to vary the entrance scattering length over a wide
range by tuning the resonance position via optical or magnetic
fields. Manipulating interactions in this way is an indispensable tool in
ultracold atomic physics and paved the way for landmark quantum
many-body experiments such as the formation of superfluid solitons
\cite{Donley2001}, the realization of the BEC-BCS crossover
\cite{PhysRevLett.92.040403, PhysRevLett.92.120403}, and self-bound
quantum droplets \cite{Cabrera301}.
\section{System under study}
Figures~\ref{fig:wavefunction}(a)--\ref{fig:wavefunction}(c) presents a scenario where a shape resonance and a
magnetic Feshbach resonance are simultaneously in play. The shape
resonance is associated with open-channel scattering-state
wave functions (blue) and manifests as an amplitude increase (dark blue) behind the
centrifugal barrier at at a particular energy [see Fig.~\ref{fig:wavefunction}(b)] The Feshbach resonance on the other hand, relates
to a bound state with a wave function (orange) which decays rapidly
with atomic separation. In Fig.~\ref{fig:wavefunction}(b) this state is
depicted to lie below the open channel threshold at $E=0$. For a
magnetic Feshbach resonance as we explore here, the closed channel hosting the bound state, has a different magnetic moment to
the open channel state.  In this two-channel setting,
\fref{fig:wavefunction}(c) shows the poles on the so-called L($-+$)
Riemann sheet \cite{Badalyan1982,Peierls1959} of the analytically
continued $S$-matrix. If coupling between the  channels is
introduced, a bound state in the closed channel with $E>0$ becomes an
``unstable bound state''\cite{Badalyan1982} that can decay into the open channel by a change in spin state. By adjusting an external magnetic field, the offset
between the open and closed channels can be varied as indicated by
the arrows in Figs.~\ref{fig:wavefunction}(b). As the bound state
moves to a positive energy, coupling to the open channel makes it
unstable and it becomes the origin of a collisional resonance while
its corresponding $S$-matrix pole moves off the real energy axis in
\fref{fig:wavefunction}(c).
\begin{figure*}[tb!]
	\centering
	\includegraphics[width=0.8\linewidth]{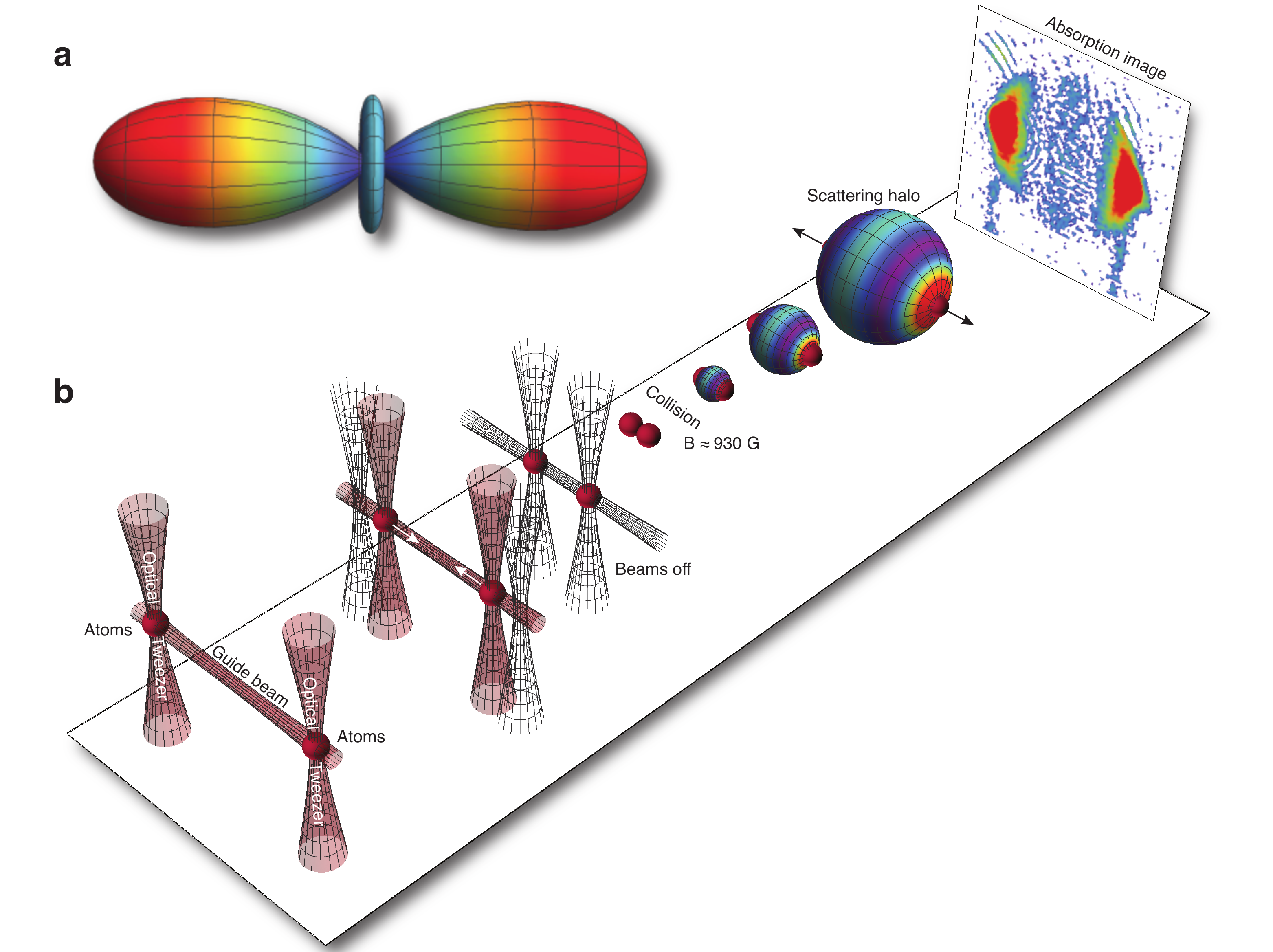}
	\caption{\label{fig:setup} { $d$-wave scattering in our
			optical collider.} {(a)} $d$-wave angular emission
		pattern which dominates the distribution of elastically
		scattered atoms in the collider experiment. {(b)} Optical
		collider procedure. Atomic clouds confined by laser beams
		are steered to collide in free space while subject to a
		magnetic field. The resulting $d$-wave dominated scattering
		halo is detected using absorption imaging.}
\end{figure*}

A theoretical description of the interaction between a shape resonance
and Feshbach resonance has been established by D\"{u}rr \textit{et
	al.}  \cite{Drr2005}, who provided a framework for interpreting
experiments on the dissociation of rubidium Feshbach molecules into
multiple partial waves. In particular, a $d$-wave shape resonance made
a compelling spatial imprint on the halo of dissociating molecules
\cite{Volz2005}. As the starting point of these experiments was a true
bound molecule, the explanation required the introduction of the
concept of ``half collisions'' as the dissociation experiment did not
include the incoming stage of a collision.  In contrast, we
investigate the interplay between a Feshbach resonance and a shape
resonance in a genuine collision experiment and seek an interpretation
in terms of an avoided crossing of scattering poles. Along with a
$d$-wave shape resonance at $E/k_B \sim \SI{300}{\micro\kelvin}$, we
make use of a narrow $d$-wave Feshbach resonance near \SI{930}{\gauss}
for ultracold ${}^{87}$Rb in the $\ket{F,m_F} \equiv \ket{1,\,1}$
hyperfine state \cite{PhysRevLett.89.283202}. This Feshbach resonance relates to a closed channel $\ell=2$ molecular state designated by the quantum numbers $(F_1=2,F_2=2)v'=-5,M=2$, where $M=m_{F_1}+m_{F_2}$ and $v'$ is the vibrational quantum number counting from the $(F_1=2,F_2=2)$ threshold. Represented in the separated atomic basis the molecular state decomposes predominantly into $|2,0\rangle+|2,2\rangle$, and $|2,0\rangle+|2,2\rangle$ (cf. Fig.\ref{fig:wavefunction} and \cite{dummy}).
\section{Experiment}
\subsection{Procedure}
Our experimental setup has been described previously
\cite{Thomas2018}. In brief, we begin our collider experiment with a
cloud of ${}^{87}$Rb atoms in the $\ket{2,2}$ state held in an optical
dipole trap. The dipole trap is formed by a crossed pair of
red-detuned laser beams---a static horizontal beam defines the
collision axis, while a vertical beam is steered by a pair of crossed
acousto-optic deflectors \cite{chisolm2018}. By rapidly toggling the
frequency of the deflector rf drive, we produce a pair of
time-averaged optical traps, which we pull apart along the horizontal
beam forming two clouds separated by \SI{0.5}{\milli\metre}. We then
transfer the cloud to the $\ket{1,1}$ state by adiabatic rapid passage
\cite{Sawyer2019} using a swept microwave field in the presence of a
small magnetic bias field. The clouds are evaporatively cooled to a
temperature of $\sim$\SI{800}{\nano\kelvin} by lowering the power of
the horizontal trap beam. To minimize heating of the clouds at our
higher collision energies, the clouds are then further separated up to
\SI{2}{\milli\metre}, providing a longer run-up and reducing the
magnitude of the acceleration needed.
\begin{figure*}[bt!]
	\centering
	\includegraphics[width=\linewidth]{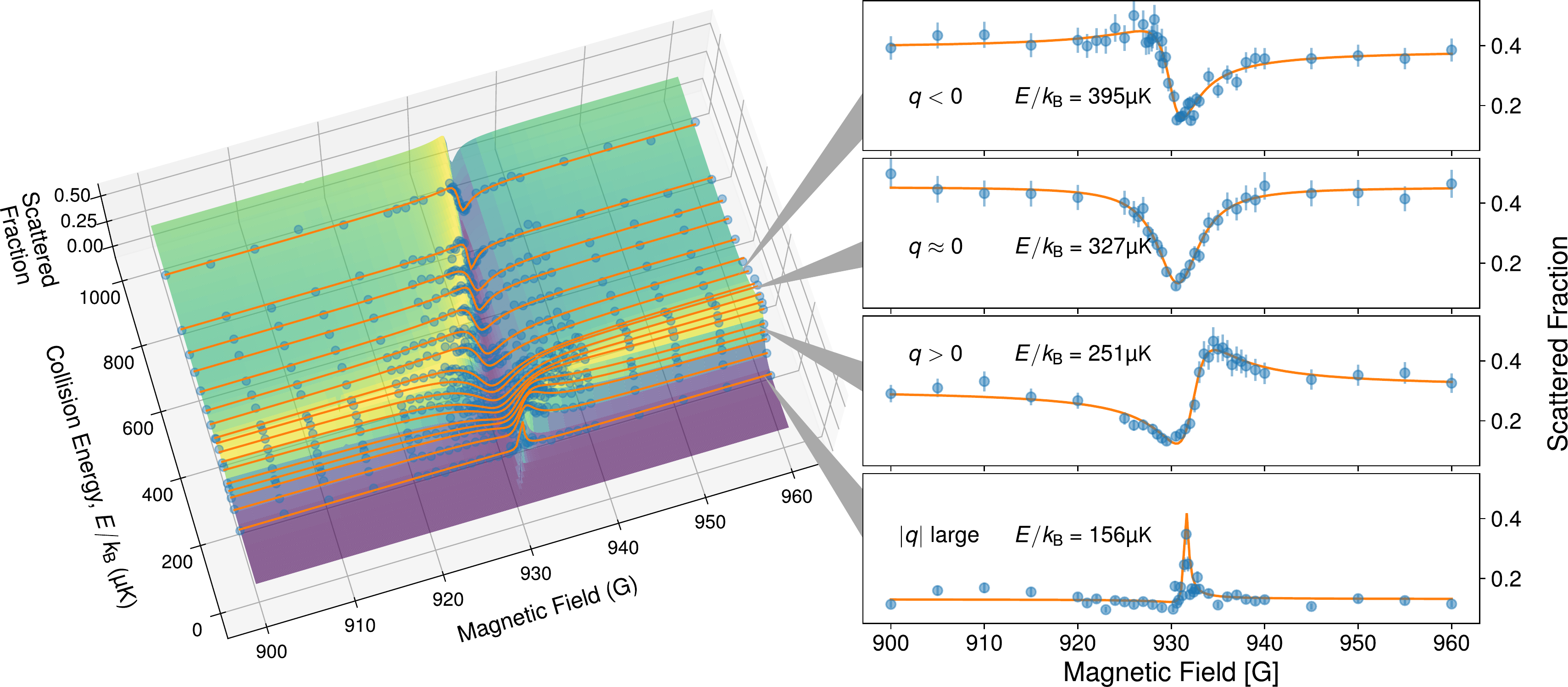}
	\caption{\label{fig:fano} {Fraction of scattered atoms as
			a function of both magnetic field and collision
			energy}. Experimentally acquired data is shown as blue
		dots while the orange lines are curve fits of Beutler-Fano
		lines shapes given by \eref{eqn:bfano}. The
		surface is an estimate of the scattered fraction based on
		coupled-channels predictions of the elastic cross section. Insets:
		Fano profiles at selected collision energies spanning the shape
		resonance, with error bars showing the standard deviation.}
\end{figure*}

Once two separated clouds have been prepared, we apply a magnetic
field along the collision axis using a pair of water-cooled coils in
the Helmholtz configuration, which carries a current regulated to the
sub-ppm level \cite{Thomas2020}. The field strength is selected to
inspect the \SI{930}{\gauss} Feshbach resonance and the vertical
trapping beams are steered to accelerate the atomic clouds towards
each other as illustrated in \fref{fig:setup}. All trapping beams
are turned off before the collision and the clouds collide in free space in
the presence of the field from the Helmholtz coils.  Following the
collision, we perform absorption imaging of the scattering halo which projects the atomic density distribution onto a plane [\fref{fig:setup}(b)]. We integrate the acquired absorption image perpendicular to the collision axis
and fit a model based on the projections of both the unscattered
clouds and the relevant scattered partial-wave components. The curve
fit extracts the fraction of scattered atoms, $\mathcal{S}$, which can
be related to the scattering cross section $\sigma$ by
\begin{equation}\label{eqn:scatt}
	\mathcal{S} (B,E) = \frac{\alpha(E)\sigma(B, E)}{1 + \alpha(E)\sigma(B, E)}.
\end{equation}
Here, $\alpha$ is a fitted parameter in our model whose value depends
on the density and cross-sectional geometry of the cloud
\cite{Thomas2018}.
\subsection{Measurements}
Figure \ref{fig:fano} shows our
measurements of $\mathcal{S}(B,E)$ over a domain spanning magnetic
fields from \SIrange{900}{960}{\gauss} and collision energies ranging
from \SIrange{150}{850}{\micro\kelvin}. We also present predictions
based on coupled-channels calculations of the elastic cross section
(See appendix \ref{app1}) which displays excellent agreement with our experimental
measurements. In the experiment, the scattered fractions were acquired
by fixing $E$ while scanning $B$, and insets in \fref{fig:fano} show
four exemplary B-field scans which capture resonance features that are
all well described by a Beutler-Fano line shape
\cite{Friedrich2013}. As the energy passes over the nominal shape
resonance position (\SI{\sim 300}{\micro\kelvin}), we observe a
handedness flip of the Fano profile (see insets of
\fref{fig:fano}). Additionally, we can see that the profile broadens
for scans around \SI{300}{\micro\kelvin} as a direct result of the
increase in the open-channel short-range wave-function amplitude [see
\fref{fig:wavefunction}(b)] enhancing the coupling of incoming
particles to the bound state in the closed channel.
\begin{figure*}[t!]
	\centering
	\includegraphics[width=0.7\linewidth]{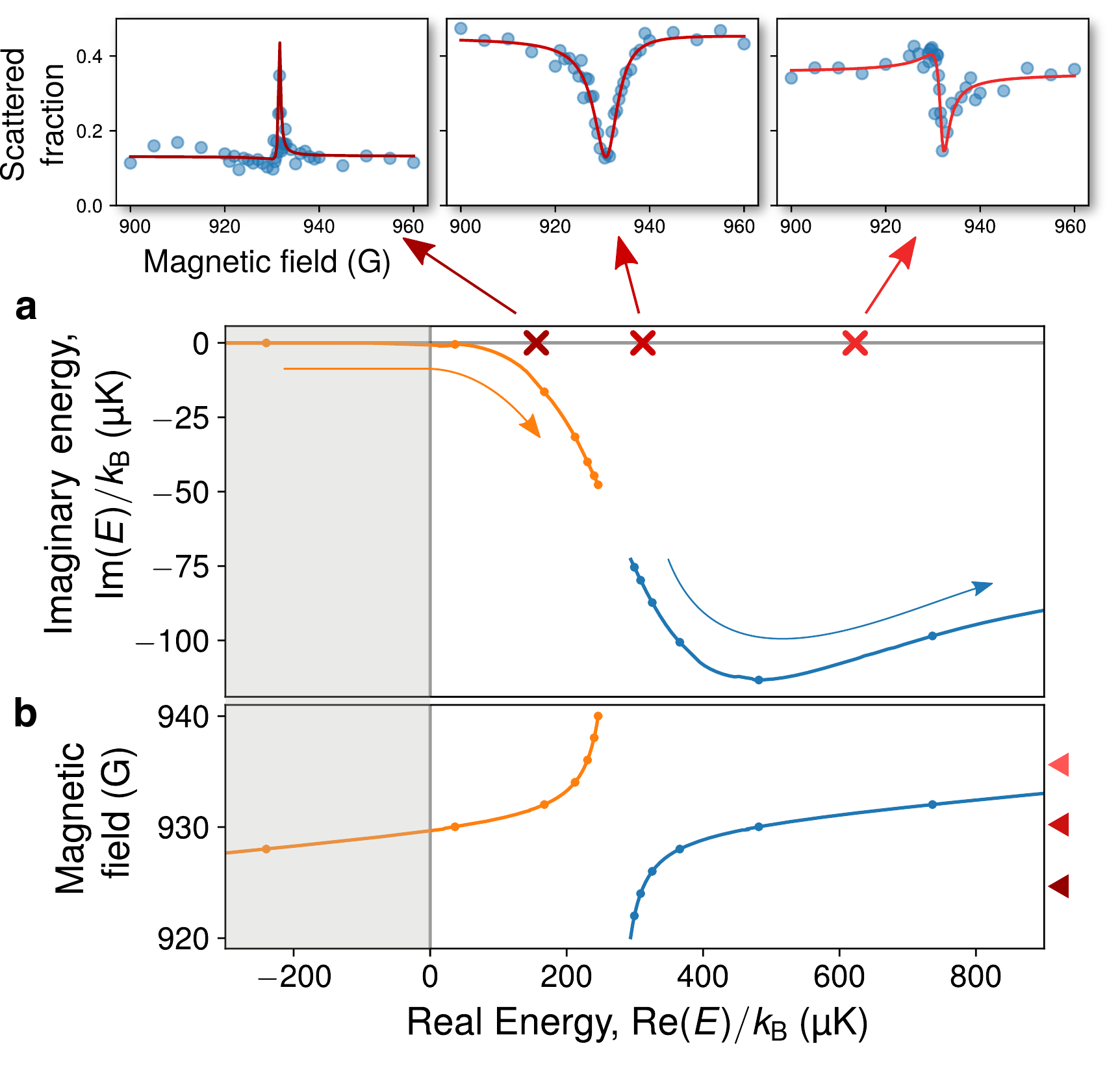}
	\caption{\label{fig:poledance}{(a)} Rb $S$-matrix pole trajectories  as
		functions of magnetic field with dots spaced every 2 G. The two poles
		can be seen to move quickly when outside of the vicinity of
		\SIrange{200}{400}{\micro\kelvin}. Inset above shows the observed
		scattering at 3 fixed energies. The narrow Fano profile
		at \SI{156}{\micro\kelvin} is a result of the orange pole moving
		past quickly. At \SI{303}{\micro\kelvin}, the broad Fano dip is
		the combined effect of the blue pole leaving, and the orange pole
		arriving and their intermediate interference. The resonance at \SI{697}{\micro\kelvin} is due
		to the pass-by of the blue pole.
		{(b)} The avoided crossing nature of the trajectories of {(a)}
		highlighted by viewing the real energy position of the
		poles in the magnetic field. The region below
		threshold is shaded grey, and the data in this region
		is derived from direct calculations of bound state
		positions in the coupled channels.}
\end{figure*}
\subsection{Beutler-Fano lines}
The Beutler-Fano line shape which describes the near-resonance
$d$-wave scattering is a function of magnetic field $B$ and collision
energy $E$. When combined with the cross section from other partial
waves, $\sigma_{\ell \neq 2}$, we have
\begin{widetext}
\begin{equation}\label{eqn:bfano}
	\sigma( E,B) = \underbrace{\frac{4\pi\hbar^2(2\ell +
			1)}{mE}\left\{\frac{[\epsilon(E,B) + q]^2}{(1+q^2)[1+\epsilon(E,B)^2]}\right\}}_{\sigma_{\ell=2}}
	+ \sigma_{\ell \neq 2}(E),
\end{equation}
\end{widetext}
with the dimensionless parameter $\epsilon(E, B) = 2(B - B_0) /
\Gamma(E)$, where $B_0$ and $\Gamma$ are the position and width of the
resonance respectively.  The parameter $q$ defines the shape of the
Fano profile, and changes sign as the profile flips in
\figref{fig:fano}---a phenomenon known as $q$-reversal \cite{Drr2005}.
We fit each data scan to a model which combines Eqs.~(\ref{eqn:scatt}) and
(\ref{eqn:bfano}) and plot the fitted curves in
\figref{fig:fano}. From the fits we extract the cross section at a slice
of the magnetic field at constant energy, along with the shape
parameter $q$ and width $\Gamma$.
\section{Interacting S-matrix poles}
To interpret our experiment within the framework of $S$-matrix
scattering poles, we analytically continue the $S$-matrix provided by
our coupled-channels calculations into the complex energy plane by
fitting a Pad\'e approximant (See appendix \ref{app2}).

In \figref{fig:poledance}(a), we plot the trajectory of the poles as
they move in the complex plane with an increasing magnetic field (see also \href{https://www.physics.otago.ac.nz/data/nk/files/AnaCont.mp4}{\textit{SI Movie 1} \cite{movielink}).}
The position of a pole, $z = E - \frac{i}{2}\gamma$, provides
the energy $E$ and width $\gamma$ of a resonance experienced along
the real energy line. The pole following the blue curve is initially the
source of the shape resonance, while the pole following the orange
curve, initially residing well below threshold, is associated with the
bound state in the closed channel, that is, the budding Feshbach resonance. As
we increase the magnetic field, both poles move to higher (real)
energies, and eventually the orange pole takes over the position of
the shape resonance, while the blue pole continues tuning as the
Feshbach resonance. In the intermediate region where the two
resonances interact, the role of each pole is not clearly defined.

Inset above the pole trajectories in \figref{fig:poledance}(a), we show the
magnetic resonance as witnessed at three fixed energies---our
collected data---and can observe which poles contribute to the form of
the resonance. In particular, we observe that at energies well below
the nominal position of the shape resonance
(\SI{~300}{\micro\kelvin}), the movement and presence of the orange
pole is responsible for the profile in magnetic field. At energies
above \SI{300}{\micro\kelvin}, it is the blue pole moving past that
makes an imprint on the scattered fraction. Close to
\SI{300}{\micro\kelvin}, both poles are involved, with their
interference resulting in a dip profile. Here, the poles swap roles
and avoid crossing in the complex energy plane, akin to the well-known
Landau-Zener avoided crossing \cite{Rotter2001}. The avoided crossing
is captured in \figref{fig:poledance}(b) which shows the real energy
of the poles during a magnetic field sweep. The two poles never
coincide at the same real energy.
\begin{table*}[]
	\centering
	\renewcommand{\arraystretch}{1.3}
	\caption{\label{tbl:pclass}Classification of $S$-matrix pole interaction}
	\begin{tabular}{p{0.24\linewidth}ccc}
		%\hline
		& Case I & Case II & Case III\\
		\hline
		$(\epsilon_1 - \epsilon_2)^2 + 4\omega^2$ & Positive & Zero & Negative\\
		%\hline
		Coupling strength $\omega$  & $ < \frac{1}{4}|\gamma_1 - \gamma_2|$ & $ = \frac{1}{4}|\gamma_1 - \gamma_2|$ & $> \frac{1}{4}|\gamma_1 - \gamma_2|$ \\
		%\hline
		\multirow{2}{*}{Crossing} &  Real energy & Poles coincide & Imaginary energy\\
		& $\real \mathcal{E}_+(B_0) = \real \mathcal{E}_-(B_0)$ & $\mathcal{E}_+(B_0) = \mathcal{E}_-(B_0)$ & $\imag \mathcal{E}_+(B_0) = \imag \mathcal{E}_-(B_0)$\\
		%\hline
		Pole trajectories given by \eref{eqn:eig} (assuming $E_1\propto B$ and $E_2$, $\gamma_1$, and $\gamma_2$ constant).
		% Trajectory for $\epsilon_1$ moving linearly past a constant $\epsn real energy (see \eref{eqn:eig})
		& \raisebox{-4.5em}{\includegraphics[width=0.22\linewidth]{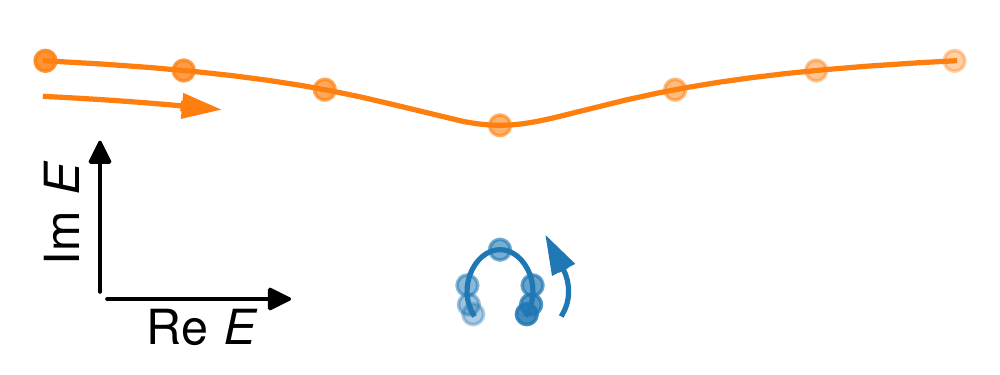}} & \raisebox{-4.5em}{\includegraphics[width=0.22\linewidth]{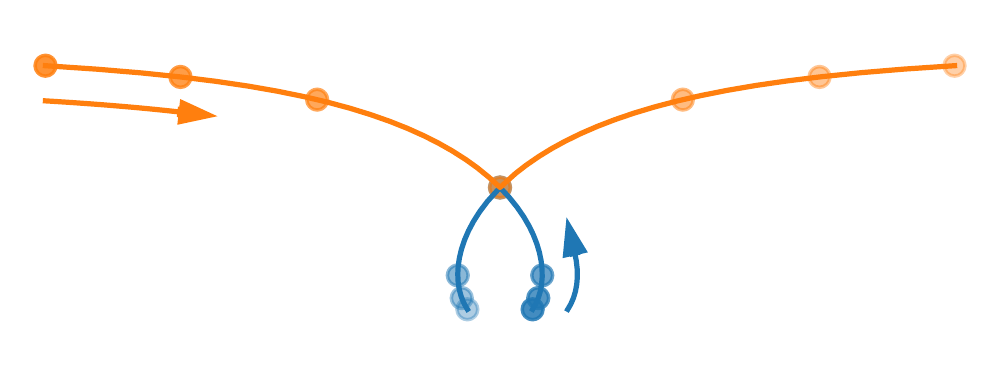}} & \raisebox{-4.5em}{\includegraphics[width=0.22\linewidth]{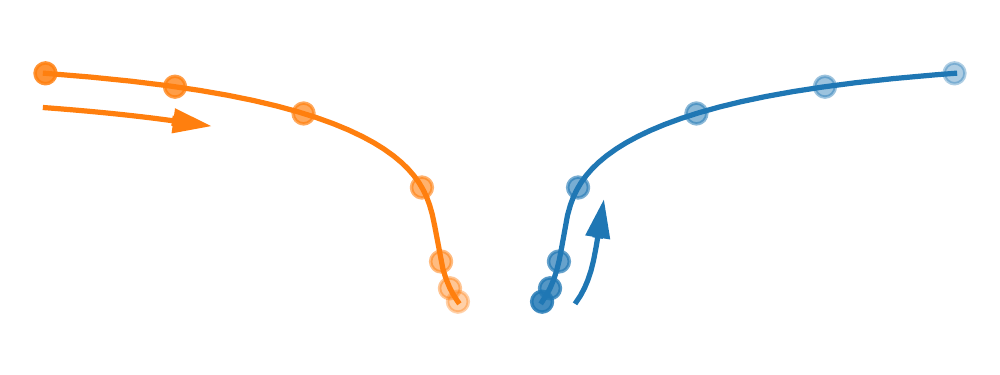}}\\
		\hline
	\end{tabular}
\end{table*}
\section{Qualitative two-level model}
The essential physics of the crossing phenomenon is revealed by
considering a coupled two-channel model described
by the effective Hamiltonian \cite{Brentano1999,Rotter2001, Okoowicz2003}
\begin{equation}\label{eqn:2ch}
	H = \begin{bmatrix}\varepsilon_1(B) & \omega \\ \omega & \varepsilon_2(B)\end{bmatrix},
\end{equation}
with inter-channel coupling $\omega > 0$ and complex bare energies
$\varepsilon_n(B) = E_n(B) - \frac{i}{2}\gamma_n$ which we assume
intersect in their real parts at $B = B_0$ such that
$E_1(B_0)=E_2(B_0)$.  The effective Hamiltonian in \eref{eqn:2ch} differs from
that of a closed two-level system by introducing an imaginary
component to the diagonal elements, representing a decay into the
continuum and making the matrix non-Hermitian. The eigenvalues of this
system, and hence the poles of the coupled $S$-matrix, are
\begin{equation}\label{eqn:eig}
	\mathcal{E}_\pm = \frac{\varepsilon_1 + \varepsilon_2}{2} \pm \frac{1}{2}
	\sqrt{(\varepsilon_1 - \varepsilon_2)^2 + 4 \omega^2}.
\end{equation}
As previously established \cite{Okoowicz2003}, the character of the avoided
crossing can be classified into three cases by the argument of the
square-root in Eq.~(\ref{eqn:eig}) at $B = B_0$ evaluating positive, zero, or negative.
This is equivalent to comparing $\omega$ to the difference in the state widths, $|\gamma_1-\gamma_2|$, as summarized in \tref{tbl:pclass}.

When $\omega=|\gamma_1-\gamma_2|/4$ (case II), the two states and the corresponding poles
will coalesce exactly ($\mathcal{E}_+=\mathcal{E}_- $) at a so-called exceptional point \cite{Kato1966,	Heiss2012} on the complex
plane for $B = B_0$. Unlike a closed two-level system, this case of a ``pole collision'' \cite{McVoy1967} displays
crossing for a non-zero coupling. In contrast, for  $\omega<|\gamma_1-\gamma_2|/4$ (case I)
the two poles will coincide only in real energy, with one
pole moving in front of the other.  Finally, for $\omega>|\gamma_1-\gamma_2|/4$  (case
III), the poles will avoid crossing in real energy, but
they will coincide in imaginary energy at $B_0$ where they represent
two equally wide resonances. In the flow of poles, one appears to
displace the other---this is exactly the strongly coupled situation established in our
collision experiment.

\section{Shape resonance shift}  A striking effect of the
avoided crossing in our experiment is a movement of the
\SI{300}{\micro\kelvin} shape resonance feature, which happens even
before the ``orange pole'' in \figref{fig:poledance} crosses the open
channel threshold.  \Fref{fig:polemove} shows how the shape
resonance moves, nearly doubling in energy as $B$ changes from
\SIrange{925}{929}{\gauss}.  This arises through strong coupling from
atomic pairs in the $d$-wave entrance channel to the closed channel
molecular bound state---the origin of the Feshbach resonance---which
has $d$-wave character. Interestingly, while the effect is dramatic at
higher energies, the scattering behaviour at threshold is not
affected.  This is a consequence of Wigner's threshold law
\cite{PhysRev.73.1002} for partial-waves with $\ell > 0$, which is
ultimately a statement of restricted tunnelling to the inner potential
region as the centrifugal barriers become wide at low
energies. Meanwhile, for our system the coupling of the entrance
$s$-wave channel and the $d$-wave bound state is small, so that
the $s$-wave scattering at threshold is not significantly affected by
the presence of the closed channel bound state. Finally,
\figref{fig:polemove} also shows how the shape resonance eventually
reemerges at its nominal position when the magnetic field is further
increased.
\section{Discussion}
\begin{figure}[b]
	\includegraphics[width=\linewidth]{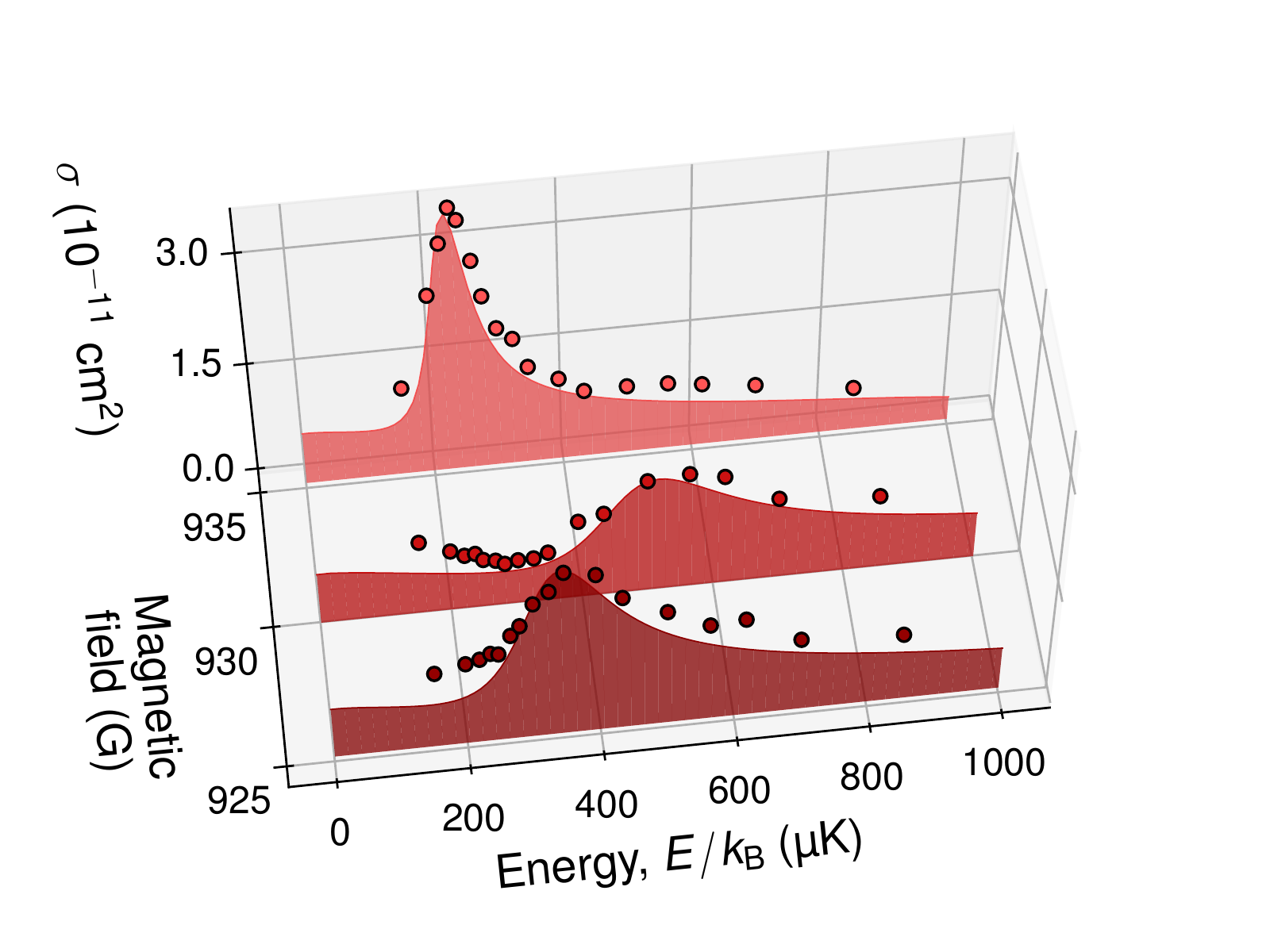}
	\caption{\label{fig:polemove} Measured and calculated cross section at three
		magnetic fields, demonstrating the movement and restoration of the
		shape resonance during the avoided crossing. The colour of the curves
		correspond to the arrows at the right of \figref{fig:poledance}B.}
\end{figure}

We have studied the interplay between two strongly interacting complex $S$-matrix poles that both reside above the open channel threshold and made use of a shape resonance and a tunable Feshbach resonance for this purpose [see \fref{fig:wavefunction}(d)].
Our results shown in Figs.~\ref{fig:fano}--\ref{fig:polemove}  are distinctly different from those obtained in the theory work by D\"{u}rr
\textit{et al.} \cite{Drr2005}, who also considered cold collisions of $\ket{1,1}$ rubidium atoms, but unlike our study focused on a Feshbach resonance at 632~G. In particular, they showed that when combining this Feshbach resonance with the shape resonance, their equivalent of our \fref{fig:polemove} describes a narrow Fano
profile performing a $q$-reversal as it moves across a stationary shape
resonance feature. Indeed, our calculations show they examined a case I interaction (see \tref{tbl:pclass}).

In a previous work \cite{Thomas2018}, we explored the interplay between
a magnetic Feshbach resonance and a virtual (antibound) state in the
collision of K and Rb. There we observed a non-monotonic
trajectory of the Feshbach resonance in a parameter space spanned by
$E$ and $B$ as predicted by Ref.~\citenum{Marcelis2004}. This result can also be interpreted in terms of interacting $S$-matrix poles. However, with the
antibound state and its corresponding $S$-matrix pole residing below threshold at all times, our collider experiments, requiring $E>0$, only allowed us to see half the picture of the associated
pole flow.

%\section{Conclusion}
The experiments presented in this article have elucidated the behaviour of two strongly coupled
$S$-matrix poles for colliding ${}^{87}$Rb atoms in an
$E$-$B$-parameter space where a shape resonance and a magnetic Feshbach
resonance are simultaneously present. While $S$-matrix poles
constitute a well-established paradigm of quantum scattering, an experimental
observation of the  avoided crossing between two such poles has so far been missing. The tunability of the  Feshbach resonance provides the crucial agent
that allows us to record the imprint of this intriguing phenomenon on
a quantum system.
Curiously, in classical systems such as coupled microwave\cite{PhysRevLett.85.2478} and optical resonators \cite{Tu:10}, the interactions of poles and their roles
in forming Fano-Feshbach resonances and exceptional points have been studied extensively, underscoring the ubiquity of these phenomena in
coupled resonant systems \cite{Joe_2006,Heiss2014,RevModPhys.82.2257, Limonov2017, Kamenetskii2018, Miri2019, Wiersig2020}.

For atomic scattering, the ``collision'' of $S$-matrix poles at an exceptional point \cite{McVoy1967} (case II in  \tref{tbl:pclass}) remains to be experimentally observed. As noted in Ref.~\citenum{Hernandez2006}, two tuning parameters are needed to achieve this in general, whereas the experiments reported in this article only has one---the external magnetic field. In the future, the additional tuning knob needed to access an exceptional point in a collision experiment might be provided by auxiliary electric \cite{Marcelis2008} or optical \cite{Bauer2009} fields.

\begin{acknowledgments}
We thank James Croft for critically reading our manuscript and Amita Deb for useful discussions.
This work was supported by the Marsden Fund of New Zealand (Contract No. UOO1923).
\end{acknowledgments}
\appendix
{\section{Coupled channels calcuations}
	\label{app1}
	To compute
	elements of the $S$-matrix, we solve the radial Schr\"odinger
	equation for the atomic separation using numerical integration. The system is
	described using a coupled-channel Hamiltonian for rubidium derived
	from a model by Strauss \textit{et al.} \cite{PhysRevA.82.052514},
	additionally including Casimir-Polder long-range interactions
	\cite{PhysRevA.59.1936}. The Hamiltonian is experimentally
	determined and models the internal magnetic structure of the atom,
	$X^1\Sigma_+$ and $a^3\Sigma^+$ molecular potentials, dipole-dipole
	interactions, and the centrifugal barrier of the $\ell > 0$ angular
	momentum states. The integration of the coupled equation uses the log-derivative method of Manolopoulos \cite{doi:10.1063/1.451472} as described in Ref.~\citenum{Thomas2017} and implemented in MATLAB computer code \cite{try}.
	\section{Analytic continuation of $S$-matrix}	
	\label{app2}
	A function over the field of complex numbers is said to be analytic if
	its complex derivative is well defined. Analytic functions are subject
	to a uniqueness theorem which states that if two analytic functions
	are identical along a curve, then they are identical everywhere. The
	$S$-matrix is analytic and in our work it is provided on the
	real, positive energy axis by our coupled channels calculations (see appendix \ref{app1}). In practice, analytical continuation into the complex plane can be perfomed by finding a so-called Pad\'e approximant that
	matches the calculated $S$-matrix on the real line.

The Pad\'e approximant \cite{Baker1996} is a rational of two finite-order
	polynomials,
	\begin{equation}
		f^{[N,M]}(z) = \frac{P(z)}{Q(z)},
	\end{equation}
	where $P(z)$ and $Q(z)$ are polynomials of degree $N$, and $M$
	respectively, given by
	\begin{subequations}
	\begin{equation}
		P(z) = \sum_{i=0}^N a_i z^i ,	\end{equation}\begin{equation}  Q(z) = \sum_{i=0}^M b_i z^i,
	\end{equation}
\end{subequations}
	with $b_0 = 1$ to fix the scaling of the
	coefficients. Pad\'e approximants can be considered a generalisation
	of the Taylor polynomial to include the presence of poles.
	
	We fit a Pad\'e approximant $f^{[4,4]}(z)$ of fourth order in both numerator and
	denominator  to the $S$-matrix calculated at a given
	magnetic field by linear least-squares methods
	\cite{Kukulin1989,Bingham2019}. From the fitted $f^{[4,4]}(z)$, the
	$S$-matrix poles are readily located by finding the roots of the
	polynomial $Q$.

%\bibliography{references_nk}
%apsrev4-2.bst 2019-01-14 (MD) hand-edited version of apsrev4-1.bst
%Control: key (0)
%Control: author (8) initials jnrlst
%Control: editor formatted (1) identically to author
%Control: production of article title (0) allowed
%Control: page (0) single
%Control: year (1) truncated
%Control: production of eprint (0) enabled
%
\end{document}